\documentclass[11pt,preprint]{aastex}

\def\iso#1#2{\mbox{${}^{#2}{\rm #1}$}}

\def\c1#1{\iso{C}{1#1}}
\def\n1#1{\iso{N}{1#1}}
\def\o1#1{\iso{O}{1#1}}

\def\gamflux{\rm photons \ cm^{-2} \ s^{-1} \ sr^{-1}}

\def\beq{\begin{equation}}
\def\eeq{\end{equation}}
\def\beqar{\begin{eqnarray}}
\def\eeqar{\end{eqnarray}}

\def\pref#1{(\ref{#1})}

\def\la{\mathrel{\mathpalette\fun <}}
\def\ga{\mathrel{\mathpalette\fun >}}
\def\fun#1#2{\lower3.6pt\vbox{\baselineskip0pt\lineskip.9pt
  \ialign{$\mathsurround=0pt#1\hfil##\hfil$\crcr#2\crcr\sim\crcr}}}

\begin{document}

\title{The Pionic Contribution to Diffuse Gamma Rays:  Upper Limits}

\author{Tijana Prodanovi\'{c}}

\and

\author{Brian D. Fields\footnotemark{$\dagger$}
\footnotetext{$\dagger$}{also Department of Physics, University of Illinois}}

\affil{Center for Theoretical Astrophysics,
Department of Astronomy, University of Illinois,
Urbana, IL 61801}

\begin{abstract}

Diffuse gamma rays probe the highest-energy processes
at the largest scales.
Here we derive model-independent constraints on the
hadronic contribution to the Galactic and
extragalactic $\gamma$-ray spectra
at in the energy range $50 \ {\rm MeV} \la E_\gamma \la 10 \ {\rm GeV}$.
The hadronic component is dominated by emission
from neutral pions, with a characteristic spectrum
symmetric about $m_{\pi^0}/2$.
We exploit the well-defined properties of the pion decay
spectrum to quantify the maximum pionic fraction of the observed $\gamma$-ray
intensity.
We find that the Galactic spectrum above 30 MeV can be at most about 50\% pionic.
The maximum pionic contribution to the extragalactic spectrum
is energy dependent; it also
depends on the redshift range over which
the sources are distributed, ranging from
as low as about 20\% for pions generated very recently,
to as much as 90\% if the pions are generated around
redshift 10.
The implications of these constraints for
models of $\gamma$-ray and neutrino emission are briefly discussed.

\end{abstract}

\keywords{cosmic rays -- gamma rays -- nuclear reactions,
nucleosynthesis, abundances}

\section{Introduction}

The prominence of diffuse emission
in the $\gamma$-ray sky above $\ga 50$ MeV
has been known since the earliest days
of $\gamma$-ray astronomy itself
\citep{fkh}.
These diffuse photons 
carry unique and direct information about some of the 
most energetic sites and processes in nature.
Diffuse $\gamma$-ray observations thus provide a
powerful tool
both (1) to test specific models of known 
or postulated astrophysical sources, and (2) to 
constrain, in a model-independent way,
known physical processes which might occur in 
one or more sources.  
We take the latter approach
in this paper, focusing in particular on
the $\gamma$-ray spectrum and 
the constraints it places on the contribution of hadronic interactions
to the overall diffuse background.

The diffuse $\gamma$-ray sky is dominated by
emission from the Galactic plane \citep{hunter},
but the presence of emission even at the Galactic poles
already suggests that an extragalactic component is
present as well \citep{sreekumar}.
The spectra of these two components
are each remarkable both for what they show
and what they do not show.
Namely, in neither spectrum is there
a strong indication of
hadronic interactions, which
are dominated by proton collisions with
interstellar matter, which
yield $\gamma$-rays predominantly through
pion production and decay:
$pp \rightarrow pp\pi^0 \rightarrow \gamma\gamma$.
The pionic spectrum is symmetric about a peak at $m_\pi/2$.
This feature, the ``pion bump,'' is notably
inconspicuous in the $\gamma$-ray data.

As we will see in detail below,
the Galactic spectrum is well-described by
a simple broken power law, with a 
break at $\sim 0.77$ GeV.
No strong pion bump is observed.
\citet{hunter} do note that
there is as a $\sim 2\sigma$ deviation in
the $60-70$ MeV energy bin, but this region in the spectrum
is otherwise well-fit by a smooth power law.
If real, this feature is remarkably narrow.
Intriguingly, detailed models of
known Galactic processes
run into difficulties explaining this
spectrum (and its simplicity).
The model of
\citet{smr00} includes
a sophisticated 2-D model of the cosmic-ray, gas,
and photon fields in the Galaxy, and includes
hadronic interactions, electron bremsstrahlung
and inverse Compton scattering of starlight.
However, when using only known cosmic ray
populations and spectra,
this model is unable to account for
the observed $\gamma$-ray spectrum.
The spectrum above about 1 GeV is flatter than
the prediction of 
pionic emission, so other sources seem to be required
as well.  Proposed explanations
for this ``GeV excess'' include
modifications to the proton spectrum,
and additional inverse Compton radiation due to
an extended halo of cosmic ray electrons
\citep{smr00}. 
One of the main goals of this paper is to quantify
the portion that can be pionic.

Information about the extragalactic component of diffuse $\gamma$-rays 
is more difficult to obtain, as one must first subtract the Galactic
foreground, which is large at low--and possibly even high--Galactic latitudes.
As we will see, the nature of the extragalactic spectrum
depends on the method used to subtract the Galactic foreground.
Different techniques have recently emerged, leading to 
different results for the shape and amplitude of the spectrum.
\citet{sreekumar} find a single power-law,
while \citet{smr03} find a smaller but ``convex'' spectrum.
In either case, no pion bump is seen.

Many astrophysical sites have been proposed to explain the
extragalactic emission.
These necessarily include ``guaranteed''
sources, namely, active \citep{ss,mc} 
and normal \citep{pf} galaxies.
These are the classes of objects which have been directly detected 
in nearby objects,
but which would be unresolved when at large distances.
Also, there is a growing consensus that the formation of large scale structures
leads to shocks in the baryonic gas, and thus to particle acceleration.
The resulting
``cosmological cosmic rays'' have recently become the subject of
intense interest \citep{mrkjco,lw,miniati,ti,kwlsh,bd,fl}.
These can also contribute to the diffuse $\gamma$-background,
and would have emission from both hadronic and inverse Compton
processes.

Here we wish to find model-independent constraints on 
hadronic and thus pionic emission mechanisms.
We choose to focus on this component because
its detection would finally confirm observationally the 
theoretical expectation that the same astrophysical acceleration processes
which give rise to non-thermal electrons (and associated synchrotron radiation)
also give rise to non-thermal ions.
Also, we wish to exploit the well-defined nature of the
pion decay spectrum which allows us to 
make a roughly model-independent comparison with observations.
Finally, since the same hadronic processes that produce
neutral pions also produced charged pions and hence
neutrinos, our limits will have implications for 
neutrino production as well.

\section{Data}
\label{sect:data}

We will consider the Galactic and extragalactic emission in turn.
For the Galactic spectrum, we adopt 
the EGRET data \citep{hunter} 
for the inner Galaxy ($300^\circ < \ell < 60^\circ$, $|b| \le 10^\circ$).
We find that the flux density can be well-fit by a broken power
law, with index $-1.52$ below $0.77$ GeV, and index $-2.25$ above: 
\beq
\label{eq:galfit}
I_{\rm obs}(\epsilon) =
  \left\{
  \begin{array}{cl}
  4.66 \times 10^{-5} \epsilon_{\rm GeV}^{-1.52} \  {\rm cm^{-2} s^{-1} sr^{-1} GeV^{-1}}
    & \epsilon_{\rm GeV} < 0.77  \\
  3.86 \times 10^{-5} \epsilon_{\rm GeV}^{-2.25} \ {\rm cm^{-2} s^{-1} sr^{-1} GeV^{-1}}
    & \epsilon_{\rm GeV} > 0.77 .  	
  \end{array}
  \right.
\eeq
This simple fit somewhat overestimates the flux in the region within about $\pm 100$ MeV
of the break, but this region will not strongly affect our
results.

Although  diffuse emission from the  Galactic plane
dominates the $\gamma$-ray sky, the emission is nonzero even at 
the Galactic poles, which suggests that there {\em is} an extragalactic component. However, it is already clear that careful subtraction will be crucial in obtaining the extragalactic gamma-ray spectrum.
Several schemes have been proposed for subtraction of the Galactic foreground. The basic approach of the EGRET team \citep{sreekumar} is to correlate the $\gamma$-ray sky with tracers of Galactic $\gamma$-ray sources. The dominant source is the hydrogen column, itself derived from observations of neutral H at 21 cm, ${\rm H}_2$ as traced by CO, and \ion{H}{2} as probed by pulsar dispersion studies. The interstellar photon field, which is up-scattered by inverse Compton processes, is also estimated. \citet{sreekumar} find evidence for a statistically significant isotropic component, with flux $I(> 100 \ {\rm MeV}) = (1.45 \pm 0.05) \times 10^{-5} \ {\rm photons \ cm^{-2} \ s^{-1} \ sr^{-1}}$ and a spectrum consistent with a single power law of index $2.1 \pm 0.03$:
\begin{equation}
\label{eq:sreek}
I_{\rm obs}= I_0 {\left( \frac{E}{E_0} \right) }^{-2.1\pm 0.03}
\end{equation}
where $E_0=0.451 \rm GeV$ and $I_0= 7.32 \times 10^{-6}\rm cm^{-2} sr^{-1} s^{-1} GeV^{-1}$.

Recently, \citet{smr03} have taken a different approach in subtracting
the Galactic foreground,
based on their sophisticated and detailed
model of the spatial and energetic content of the Galaxy.
They used the GALPROP model for cosmic ray propagation to predict the Galactic  component and give  the new estimate of the extragalactic gamma-ray background (hereafter EGRB) from EGRET data. \citet{smr03} also find evidence for an EGRB,
but with a different spectral shape, and in general a lower amplitude
than that of  \citet{sreekumar}.
The \citet{smr03} Galactic foreground estimates also includes
the \citet{smr00} estimate of the pionic contribution.
This model-based constraint will serve as an important consistency 
check of our model-independent
results.
We used the least square method to fit their data with a cubic logarithmic function for the energy range  0.05-10 GeV:
\begin{equation}
\label{eq:strong}
\ln (I_{\rm obs}E^2)=-13.9357-0.0327\ln E+0.1091(\ln E)^2+0.0101(\ln E)^3
\end{equation}
In this fit energy E is understood to be in the units of GeV
and I in units of $\gamflux \ {\rm GeV^{-1}}$.

Indeed, the latest analysis of EGRET data done by \citet{kwl} also implies that Galactic foreground was overestimated in previous work. 
They find that Galactic foreground in fact dominates the sky and that only an upper limit on the EGRB can be placed.
However, \citet{kwl} analysis did not contain spectral information which is why it is not further investigated in this paper.
The data used in this paper along with the fits are shown in Fig.\ref{fig:mw} (Galactic component) and Fig.\ref{fig:fluxes} (EGRB).

\section{A Simple Model for Pionic Gamma-Rays}

The general expression for the $\gamma$-ray intensity spectrum
at energy $\epsilon$
in a particular direction is given by the line-of-sight integral
\beq
\label{eq:flux}
I(\epsilon) = \frac{1}{4\pi} \int_{\rm los} q(\epsilon,\vec{r}) \ ds 
  = \frac{1}{4\pi} \int_{\rm los} \Gamma(\epsilon) n_H(\vec{r}) \ ds 
\eeq
where we have ignored absorption and scattering
processes which are negligible for $\epsilon \la 20$ GeV
\citep[e.g.,][]{mp,ss98}.  
In eq.\ (\ref{eq:flux}) we write the $\gamma$-ray emissivity
(production rate per unit volume) in terms of the local
hydrogen density $n_H$ and the 
production rate per H-atom \citep[][e.g.,]{stecker70,dermer}
\beq
\label{eq:source}
\Gamma(\epsilon) = \int_{\epsilon+m_\pi^2/4\epsilon}^\infty
                    \frac{dE_\pi}{\sqrt{E_\pi^2-m_\pi^2}}
                     \int dE_p \phi(E_p) \frac{d\sigma(E_p,E_\pi)}{dE_\pi}  
\eeq
Note that if the {\em shape} of the cosmic ray spectrum $\phi(E)$
is the same everywhere along the line of sight, then $I(\epsilon) = \Gamma(\epsilon) N_H$,
where $N_H$ is the hydrogen column density, and thus
the shape of the  observed $\gamma$-ray spectrum $I(\epsilon)$ is the same as
that of the source $\Gamma(\epsilon)$.  This is the case of interest to us.

The production rate $\Gamma$ reflects the production 
and decay of neutral pions (with cross section $\sigma$)
due to a cosmic ray flux spectrum $\phi$.
The shape of $\Gamma(\epsilon)$ has
well-known properties that reflect the symmetry of
the decay photons in the pion rest frame.
As described in detail by \citet{stecker70,stecker71},
the underlying isotropic nature of the rest-frame emission and 
the cosmic-ray beam is encoded in the emissivity spectrum,
whose only photon energy dependence is through the
lower limit in eq.\ (\ref{eq:source}).
This can be written as $\epsilon_0(\epsilon/\epsilon_0 + \epsilon_0/\epsilon)$
which clearly has a minimum
at $\epsilon_0 = m_\pi/2$, and is invariant under
$\epsilon/\epsilon_0 \rightarrow \epsilon_0/\epsilon$;
these properties guarantee that the spectrum is
peaked at $\epsilon_0 = 69$ MeV (the pion bump) and falls off symmetrically on
a $\log F-\log \epsilon$ plot.

The other key property of emissivity is found
in the isobar+fireball model, which provides a good
fit to accelerator data \citep{dermer}.
Namely, at high energies $\epsilon \gg m_\pi/2$, the emissivity
goes to the power law $\Gamma(\epsilon) \sim \epsilon^{-\alpha_p}$
(and thus by symmetry it goes  at low energies as $\epsilon^{+\alpha_p}$).
This simple asymptotic power-law dependence is what allows us to 
constrain the pionic contribution of $\gamma$-ray spectra.

Note that the region of the spectrum immediately around the pion bump
depends most sensitively on the details of the pion production cross
section $d\sigma(E_p,E_\pi)/dE_\pi$ and thus on the shape of
the proton spectrum $\phi_p(E)$ with which it is convolved.
Consequently, a detection of the pion bump, and its width,
would not only unambiguously identify a hadronic source,
but would also constrain the  spectrum of source particles.
In this case, our constraints, which are based on the absence of
a pion bump and the asymptotic behavior of the pion spectrum, become superfluous.
We look forward to this obsolescence, due to the eventual detection of 
the pion bump by GLAST or its successors. But until then
our results remain relevant.

A convenient semi-analytic fit to the pionic $\gamma$-ray source-function 
was recently presented by \citet{ensslin}.
Using Dermer's model \citep{dermer} for the production cross section,
they arrive at the form:
\beq
\Gamma(\epsilon) 
  = \xi^{2-\alpha_{\gamma}} \frac{n(r)_{p,\rm CR}}{\rm GeV} 
    \frac{4}{3\alpha_{\gamma}} \left( \frac{m_{\pi^0}}{\rm GeV} \right)^{-\alpha_{\gamma}} 
  \left[ \left( \frac{2\epsilon}{m_{\pi^0}} \right)^{\delta_{\gamma}} 
 + \left( \frac{2\epsilon}{m_{\pi^0}} \right)^{-\delta_{\gamma}} \right]^{-\alpha_{\gamma}/\delta_{\gamma}}
  \sigma_{pp}
\label{eq:pion}
\eeq
The spectral index $\alpha_{\gamma}$ determines the shape parameter $\delta_{\gamma}=0.14 \alpha_{\gamma} ^{-1.6}+0.44$.
The effective cross section  $\sigma_{pp}$  they modeled to the form $\sigma_{pp}= 32 \times (0.96+e^{4.4-2.4 \alpha_{\gamma}})$ mbarn. 
Following \citet{dermer} we take the pion multiplicity to be $\xi =2$. 
The cosmic ray projectile number density is $n_p(r)$. 
This source function peaks at half the pion rest energy. 
In Dermer's model the $\gamma$-ray spectral index $\alpha_{\gamma}$ 
is equivalent to the cosmic ray spectral index i.e. $\alpha_{\gamma}=\alpha_p$ \citep{dermer}.
Note that in our limits on the dimensionless {\em fraction} 
of observed emission that is due to pion decay,
only the energy dependence (i.e., the {\em shape}) of the emissivity
in eq.\ (\ref{eq:pion}) is important.

For the case of extragalactic emission,
these pionic $\gamma$-rays can come from different redshifts.
Thus, for extragalactic origin   eq.\ \pref{eq:flux} becomes
\beq
\label{eq:cosmo}
I(\epsilon) 
  = \frac{1}{H_0} \int dz \frac{n_{\rm H,com}(z)\Gamma[(1+z)\epsilon,z]}{(1+z){\cal H}(z)}
\eeq
where
the dimensionless expansion rate ${\cal H}(z)= H(z)/H_0$ takes
the form ${\cal H}(z) = \sqrt{\Omega_{\rm m} (1+z)^3+\Omega_\Lambda}$ 
in a flat universe.
The redshift dependence of the source-function $\Gamma$
depends on the nature of the emission site (galaxies, cosmological shocks, etc.).
For purposes of illustration, we will use a 
single-redshift approx $n(z) = n_0 \delta(z-z_*)$.
In this approximation different $z_*$ amount to the shift of the pionic $\gamma$-ray flux in log-log plot. Thus in this simplistic view the form of the source-function would stay the same as in equation(\ref{eq:pion}) but $\epsilon_{\gamma}$ would be substituted with $E_{\gamma}(1+z)$ where $E_{\gamma}$ is now the observed gamma-ray energy.
Note that in this case, any pion bump would be redshifted,
and thus would appear at energies $< m_{\pi^0}/2$.
Thus it is clear that this feature is not apparent
in the extragalactic spectrum, which is flat or even convex
at these energies.

Of course, any realistic case will include contributions
from a range of redshifts.  However, one can view this distribution
as an ensemble of delta functions,
which will be an averaging over our simple cases,
with a redshift-dependent weighting which scales as
$(1+z)^{-1} n_{\rm H,com}(z) {\cal H}(z)^{-1}$ (c.f.\ eq.\ \ref{eq:cosmo}).

\section{Procedure}
\label{sect:procedure}

The main goal of this paper is to place a constraint to the maximal pionic contribution to diffuse gamma-ray flux based on the shape of the pionic spectrum, 
and fact that the pion bump 
is not observed. 
The way to obtain this upper limit is to see how much can we increase the pionic 
contribution by changing the parameters that it depends on so that it always stays 
at or below the observed values at all energies. 
The parameters that we change are the projectile and target number densities that enter in cosmic ray production of pions and the redshift from where we assume all pionic gamma rays originate.
The condition of matching logarithmic slopes  
\beq
\label{eq:match}
\frac{d \log I_{\rm obs}(E)}{dE} = \frac{d \log I_{\pi^0}(E)}{dE}
\eeq 
of theoretical pionic gamma-ray flux and the fit to the observed gamma-ray flux 
guarantees that the ratio $I_{\pi^0}/I_{\rm obs}$ is extremized (and in fact
maximized for the spectra we consider). Here $I_{\pi^0}(E)=n_H(\vec{r}) \Gamma(E)$ 
and is given in units of $\rm GeV^{-1} \, s^{-1} \, cm^{-2}$.
The energy which satisfies eq.\ (\ref{eq:match}) 
thus sets the values of our parameters that maximize pionic flux.

Since the energy of pionic gamma-rays depends on the redshift as stated in the previous section, the slope of this theoretical flux will be the following function of observed energy $E$ and the redshift $z$:
\begin{equation}
\frac{d\log I_{\pi^0}}{d\log E} = -\alpha_{\gamma} \frac
  {  \left( 2E(1+z)/m_{\pi^0} \right)^{\delta_{\gamma}}
   - \left( 2E(1+z)/m_{\pi^0} \right)^{-\delta_{\gamma}} }
  {  \left( 2E(1+z)/m_{\pi^0} \right)^{\delta_{\gamma}}
   + \left( 2E(1+z)/m_{\pi^0} \right)^{-\delta_{\gamma}} }
\label{eq:pislope}
\end{equation}
Of course, for the Galactic spectrum we take $z=0$.

The choice of $\alpha_{\gamma}$ depends on the origin of cosmic rays.
In the case of Galactic cosmic rays we will be using the 
classic observed--i.e., {\em propagated}--value $\alpha_{\gamma}=2.75$ 
\citep[confirmed recently by, e.g.,][]{boezio,ams,sanuki}.
For extragalactic $\gamma$-rays,
the sources are not known, but
both blazars and shocks in cosmological
structure formation have received considerable attention. 
For the case of blazars, it is not clear
whether the emission is due to hadronic
or leptonic processes.
Blazar $\gamma$-ray spectral indices have a distribution 
which averages give to a diffuse flux with index 
$\alpha_\gamma \sim 2.2$  \cite{ss}; if the emission is
pionic this would be the proton index as well. 
Also, 
it is well known that the spectral index of cosmic rays accelerated in fairly strong shocks is $\alpha \approx -2$ \citep{be,je} which is expected to be the case with the cosmic rays from
structure formation.
Although the spectrum of structure-formation cosmic rays 
is not very well known for this purpose we will adopt the value $\alpha_{\gamma}=2.2$,
which is near the strong-shock limiting value of 2,
and consistent with the Galactic {\em source} value
\citep[see discussion in, e.g.,][]{focv},
as well as that of blazars.

Now we have to match the slopes of the observed gamma-ray spectra  
to the slope of the theoretical pionic flux that was given in equation (\ref{eq:pislope}).
This amounts to equating (\ref{eq:pislope}) with to the appropriate expressions
for the spectra:  eqs.\ (\ref{eq:sreekumarslope}) or (\ref{eq:strongslope}) for 
extragalactic, and eq.\ (\ref{eq:galacticslope}) for Galactic.
We then solve for $E_{\gamma}(z)$, 
where we put $z=0$ for the Galactic case, and $z=z_*$ for the extragalactic case.

\section{Results}

\subsection{Galactic Spectrum}

As described in \S\ref{sect:data}, we fit the EGRET data for the Galactic spectrum
with a broken power-law (eq.\ \ref{eq:galfit}),
and we use the emissivity for 
a proton spectrum $\alpha_p = \alpha_{\gamma}=2.75$.
In order to set up an upper limit to the pionic contribution we match the 
low-energy index $-1.52$ to the slope of pionic 
$\gamma$-rays; fitting to the higher energy portion  of the spectrum
would lead to an unobserved excess in the low-energy portion.
The logarithmic slope of galactic spectrum is then just
\begin{equation}
\frac{d\log I_{\rm obs}}{d\log E}=-1.52
\label{eq:galacticslope}
\end{equation}

\begin{figure}[htb]
\epsscale{0.85}
\plotone{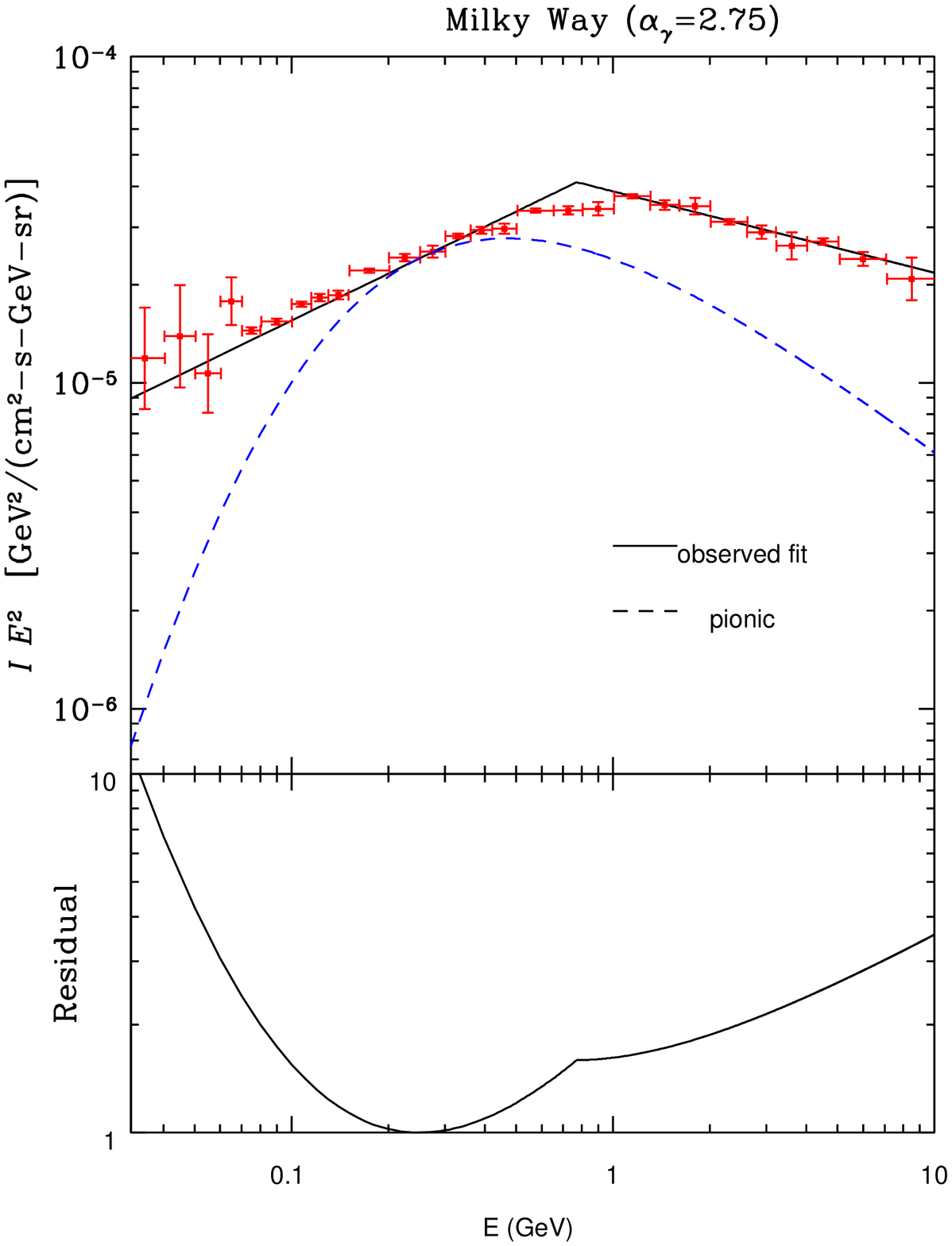}
\caption{In this figure we present the maximal pionic contribution to the 
Galactic $\gamma$-ray spectrum. EGRET data points are taken from \citet{hunter}.
The lower panels represent the residual, that is, 
$\log[(FE^2)_{\rm obs}/(FE^2)_{\pi^0}] = \log (I_{\rm \rm obs}/I_{\pi^0})$.
Note that the kink at $0.77$ GeV is unphysical and just due to the
overshooting of the simple broken power-law fit.}
\label{fig:mw}
\end{figure}

We now equate this with pionic slope qiven in eq. (\ref{eq:pislope}) and solve for $E_{\gamma}(z=0)$. This sets up the maximal  normalization to the pionic spectrum which is plotted in the Fig.\ref{fig:mw} along with the observed Galactic spectrum. Also plotted is the logarithmic residual function.
After integration over energies up to 10 GeV we can finally obtain the maximal pionic fraction of the Galactic $\gamma$-ray flux based on the shape of the pion decay spectrum as well as the lack of as strong detection of the pion bump:
\beq
f_{\pi^0,\rm MW}(>\epsilon) 
 = \frac{I_{\pi^0,\rm max}(>\epsilon)}{I_{\rm \rm obs}(>\epsilon)}
\eeq
where $I(>\epsilon)=\int_{\epsilon} I(E)dE$.
We find pionic fraction to be $f_{\pi^0,\rm MW}(>30 \rm MeV)=53\%$ and $f_{\pi^0,\rm MW}(>200 \rm MeV)=81\% $.
While this integral constraint provides
a diagnostic of the hadronic ``photon budget,'' we stress that the
lesson of the residual plot in Fig.\ \ref{fig:mw}
is that the deficit is not at all uniform across energies, but
is very large at both high and low energies.

\subsection{Extragalactic Spectrum}

By going through the slope-matching procedure described in the previous section we can fix 
the parameters that maximize the pionic contribution to 
the different extragalactic $\gamma$-ray spectra we consider. 
For the \citet{sreekumar} spectrum (eq.\ \ref{eq:sreek}),
the logarithmic slope is just a constant
\begin{equation}
\frac{d\log I_{\rm obs}}{d\log E}=-2.1
\label{eq:sreekumarslope}
\end{equation}
On the other hand, for \citep[eq.\ \ref{eq:strong}]{smr03},
we have
\begin{eqnarray}
\frac{d\ln I_{\rm obs}}{d\ln E} &=& -2+ \frac{d\ln (I_{\rm obs}E^2)}{d\ln E} \\
&=& -2.0327+0.2182\ln E+0.0305(\ln E)^2  \label{eq:strongslope}
\end{eqnarray}

\begin{figure}[htb]
\epsscale{1.0}
\plottwo{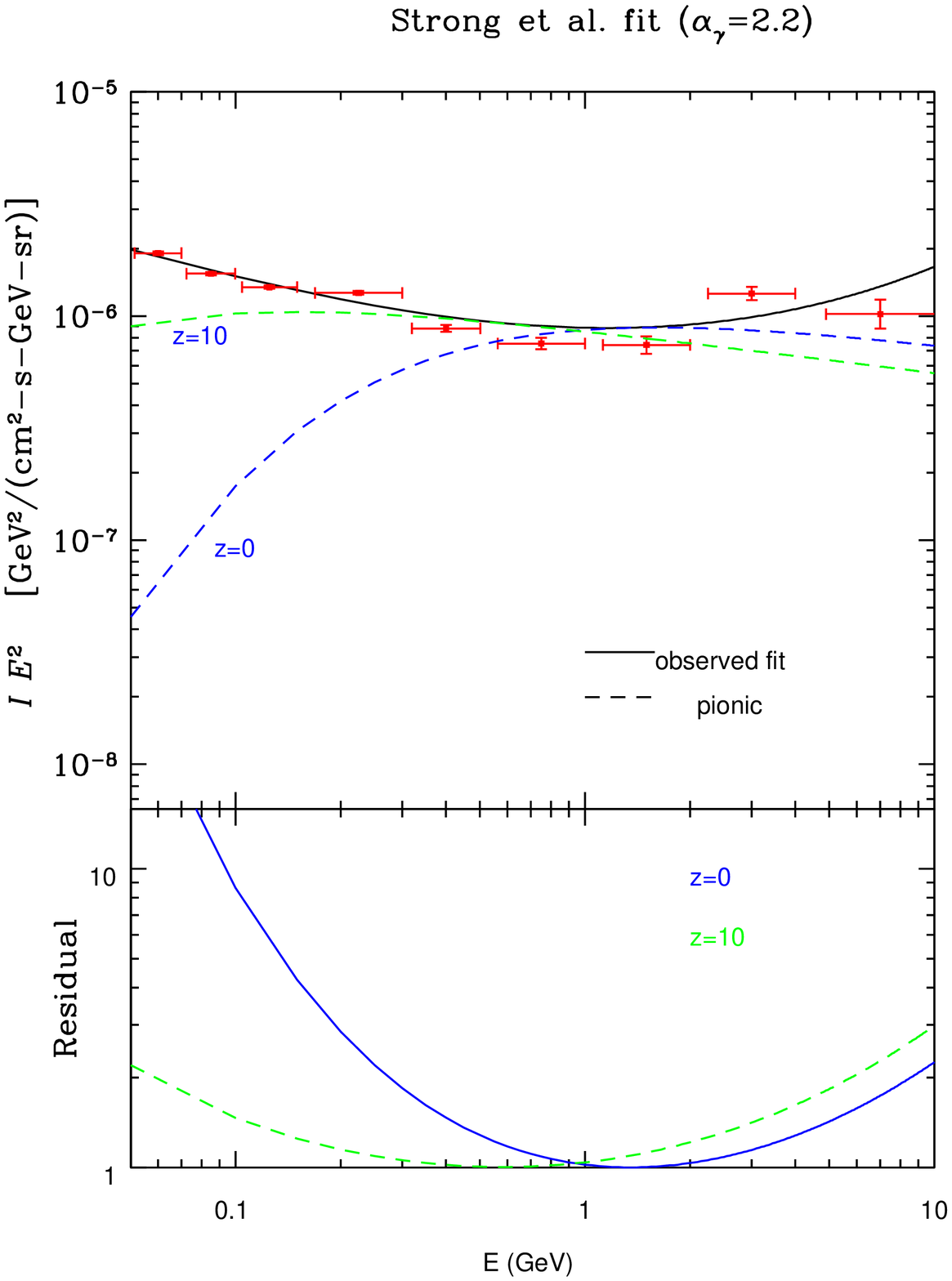}{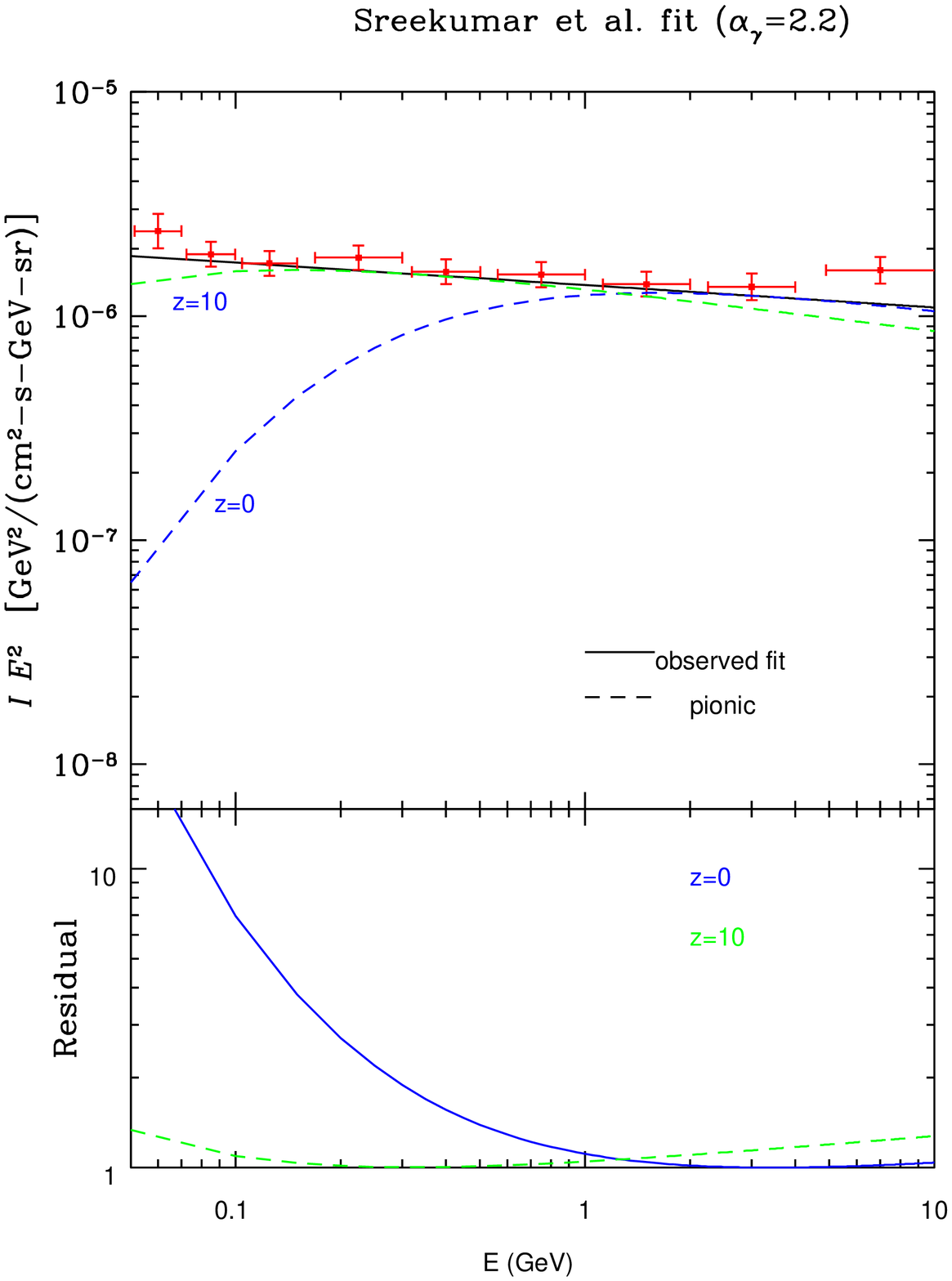}
\caption{The maximal pionic contribution to the extragalactic $\gamma$-ray spectrum,
computed by assuming that pionic $\gamma$-rays originated at a single redshift, namely at $z_*=0$ and $z_*=10$. EGRET data points for both fits taken from \citet{smr03}. Lower panels represent the residual function as in Fig.\ \ref{fig:mw}.}
\label{fig:fluxes}
\end{figure}

In our simplistic picture we assume that all of the pionic $\gamma$-rays originated at a single redshift.
Thus we go through this procedure for a set of redshifts ranging from $z=0$ up to $z=10$. 
Figure~\ref{fig:fluxes} 
shows our maximized pionic contribution for the two extreme redshifts, along with the fits to the observed $\gamma$-ray spectrum and the actual EGRET data points \citep{smr03}. We also present the residual, which is what it is left after pionic flux contribution is subtracted from the observed $\gamma$-ray spectrum.
Here we see that for both EGRB spectra, 
the residual is large at low energies. 
However, the different shapes of the two EGRB candidate spectra
lead to qualitatively different behavior at high energies ($\ga 1$ GeV):
the residual remains substantial ($\ga$ a factor of 2) for the 
\citet{smr03} fit, suggesting the need for other component(s)
to dominate both high and low energies. 
But for the \citet{sreekumar} fit, the residual is small, and thus the
pionic contribution can be dominant above 
1 GeV.  
This difference highlights the current uncertainty of our knowledge
of the EGRB spectrum \citep[and even its existence,][]{kwl}.
Our analysis thus underscores the need for a secure determination of
the Galactic foreground and the extragalactic background.

\begin{figure}[htb]
\epsscale{0.85}
\plotone{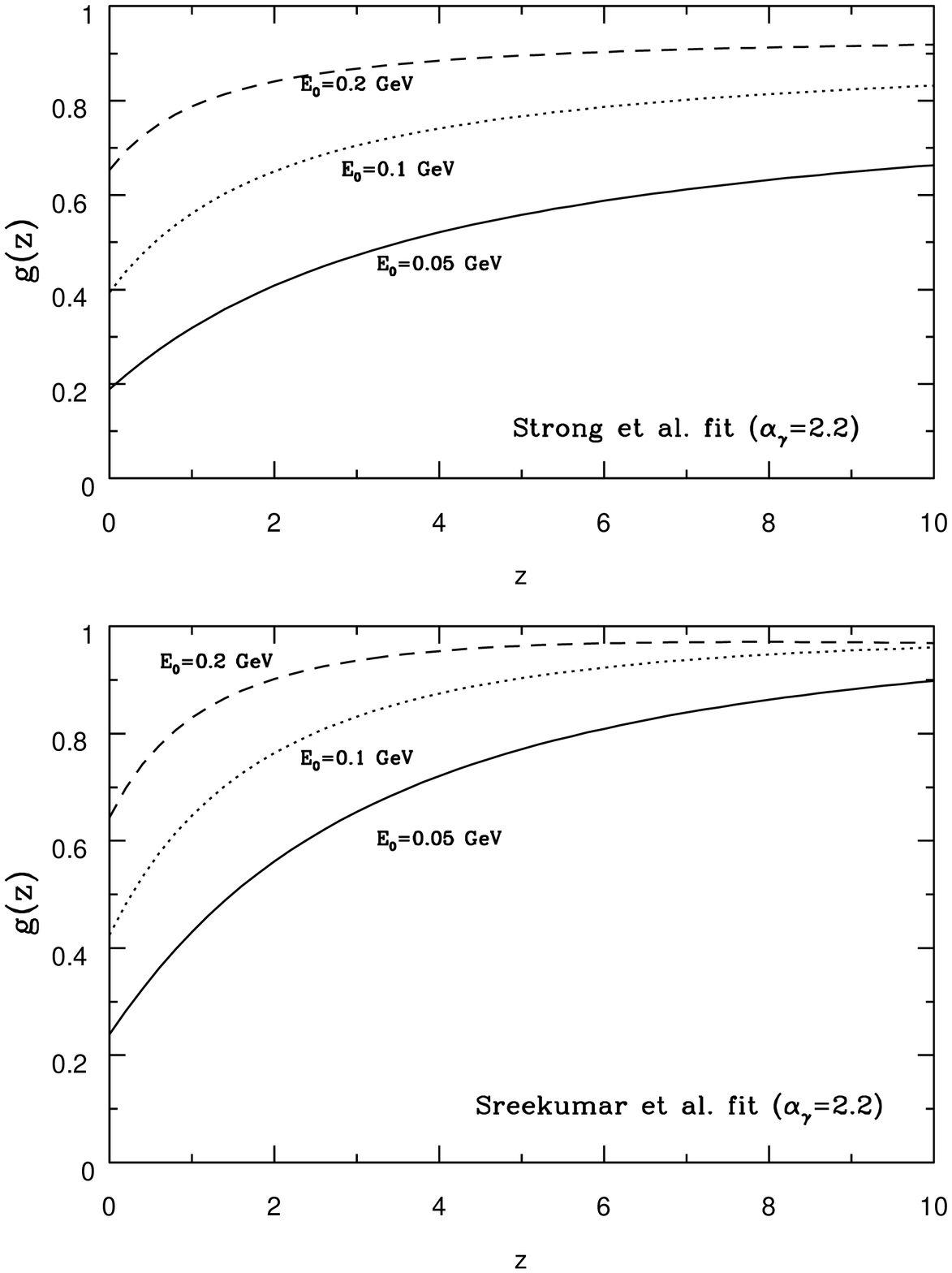}
\caption{In this figure we see the maximal fraction of pionic energy-integrated flux. It is given as a function of the redshift of origin for the pionic $\gamma$-rays. Fluxes were integrated from $E_0$ up to 10 GeV.}
\label{fig:fraction}
\end{figure}

To finally obtain the upper limit for the $\gamma$-rays that originated from pion decay, we integrate pionic and the  observed (for both fits) flux. Then the ratio
of these energy-integrated fluxes is the maximal fraction of pionic $\gamma$-rays for a given redshift.
\begin{equation}
g(z)=\frac{\int_{E_0}^{10 \rm GeV} d\epsilon I_{\pi}(\epsilon,z)}{\int_{E_0}^{10 \rm GeV} d\epsilon I_{\rm obs}(\epsilon)}
\end{equation}
In Fig.~\ref{fig:fraction} we plot this ratio as a function of redshift for three different integration ranges and for both \citet{smr03} and \citet{sreekumar} fits to EGRET data.
Note that the results asymptotically approach unity.
A glance at Figure \ref{fig:fluxes} suggests the reason for this:
the effect of increasing the emission redshift $z_*$ to 
``slide'' the pionic spectrum leftward, toward lower
energies.  As a result, the peak and low-energy falloff are redshifted out of
the fit regime, and the remaining high-energy power-law tail of the pionic emission
then provides a good fit to the observations.

\section{Discussion} 

We have presented model-independent upper limits on
hadronic $\gamma$-ray emission based on the shape of observed 
spectra and their lack of a pion bump.
Above 100 MeV, one might expect that gamma-rays from $\pi^0$ decay should dominate the Galactic spectrum.
However, we find that they can make only about 50\% of total Galactic gamma-ray flux. 
From the shape of the residual function that is plotted in Fig.\ref{fig:mw} we can see that due to the break at 0.77 GeV, additional gamma-ray components are needed
both above and below the $\sim 250$ MeV regime at which the pionic contribution
can be maximized. 
The residual gamma-rays below this scale 
can possibly be accounted by bremsstrahlung and inverse Compton scattering
\citep{smr00}.
However, if the pionic component is near its maximum, 
there still is the need for at one more component  of gamma-rays
above $\sim 300$ MeV
to account for the EGRET data; alternatively,
a single process may be at work, but with pionic emission being sub-dominant
at all energies. 
This model-independent result is in good agreement
with the sophisticated analysis of different models by \citet{smr00}.
We also stress again that this analysis was based on the assumption that the pion bump is not observed. If a future $\gamma$-ray mission such as
GLAST were to identify this feature that would allow us to set a more definite and stronger limit on the pionic fraction of diffuse gamma-rays. 

The maximum pionic fraction of extragalactic $\gamma$-rays
can be seen in Fig.\ref{fig:fraction}
for different methods of foreground subtraction. 
It can go from $20-90\%$, depending on the assumed redshift of 
cosmic ray origin,
but also on method used in subtracting the Galactic foreground
to obtain the EGRB. 
Namely, the \citet{smr03} fit gives the fraction of  20\% 
for cosmic rays that originated at the present, 
up to about 70\% for $z=10$, and in both cases
there is still a factor of $\ga 2$ deficit of high-energy
photons.
On the other hand,
the \citet{sreekumar} fit implies a 20\% pionic fraction for recent cosmic rays and about 90\% for redshift 10 cosmic rays,
with no significant deficit at high energies.. 
This large variation underscores the need for a robust procedure
for determining the EGRB, and also emphasizes the power of a firm EGRB
spectrum to constrain emission processes.

Our limits can be compared to the results of specific models.
\citet{miniati} finds that the pionic component contributes
about 30\% of the total 
emission from structure-formation cosmic rays, with the balance arising
from electron synchrotron emission.  He in turn finds that the
entire cosmic-ray component
itself can be $\sim 20-30\%$ of the total \citep{sreekumar} observed
background.  Thus, the \citet{miniati} model finds that the pionic contribution
is $\sim 6-10\%$ of the observed level, a fraction which
would be larger if the smaller \citet{smr03} background were used.
The work of \citet{kwlsh} neglects pionic emission entirely,
arguing that the synchrotron emission should dominate.

Of course, our extragalactic constraints reflect our simple
single-redshift approximation for the origin of cosmic rays. More
precise analysis should include some kind of averaging over the
redshifts. 
Further progress first requires a detailed knowledge of the redshift
evolution of sources and targets.
On the other hand, if the ``pionic bump'' was observed in
the EGRB spectrum by GLAST or other future experiments then the
position of its peak could immediately tell us something about the mean
cosmic-ray flux. This information would then give us a better handle
on the star-formation rate as a function of redshift.

Finally, we note that our constraints on photons of hadronic origin
also have implications for neutrinos produced in the same processes,
a connection which has been emphasized by \citet{wb}.
This is of particular interest in the case of extragalactic neutrino
emission, which may lead to high-energy ($E_\nu \ga 1$ TeV) events
observable by ICECUBE \citep{icecube}.
\citet{wb} use the energy density of ultra-high-energy 
($\ga 10^{19}$ eV) cosmic rays
to derive limits on the high-energy cosmic neutrino flux.
Furthermore, they find that the EGRB (in the power-law
form of \citet{sreekumar}) implies a neutrino flux above $\sim $ TeV
which violates this limit by a factor of up to $\sim 100$, if the EGRB is of hadronic origin.
Our limits on the pionic fraction of the EGRB are
complementary to the \citet{wb} result. 
Our constraints on the hadronic origin of the EGRB
are weaker than their $\sim 1\%$ fraction, but 
are derived independently, based on the EGRB spectrum
itself.

\acknowledgments
We thank Andy Strong for valuable discussions,
and we are indebted to Stan Hunter for kindly providing us with the 
EGRET Galactic spectrum.
We are grateful to Vasiliki Pavlidou for valuable 
discussions, particularly concerning neutrino production.
This material is based upon work supported by the National Science
Foundation under Grant No. AST-0092939.

\end{document}